\documentclass[reprint, amsmath,amssymb, aps,pra]{revtex4-2}
\usepackage{graphicx}
\usepackage{dcolumn}
\usepackage{xcolor}
\usepackage{physics}
\usepackage[colorlinks,linkcolor=red,citecolor=blue,urlcolor=red]{hyperref}
\usepackage{float}
\usepackage[english]{babel}

\DeclareUnicodeCharacter{0308}{\"}

\renewcommand{\t}{\tau}

\renewcommand{\t}[1]{\mathrm{#1}}

\begin{document}


\title{Membrane-based Optomechanical Accelerometry}


\author{Mitul Dey Chowdhury}    
\author{Aman R. Agrawal}
\author{Dalziel J. Wilson}
\affiliation{Wyant College of Optical Sciences, University of Arizona, Tucson, AZ 85721, USA}

\date{\today}


\begin{abstract}

Optomechanical accelerometers promise quantum-limited readout, high detection bandwidth, self-calibration, and radiation pressure stabilization. We present a simple, scalable platform that enables these benefits with nano-$g$ sensitivity at acoustic frequencies, based on a pair of vertically integrated Si$_3$N$_4$ membranes with different stiffnesses, forming an optical cavity. As a demonstration, we integrate an ultrahigh-Q ($>10^7$), millimeter-scale Si$_3$N$_4$ trampoline membrane above an unpatterned membrane on the same Si chip, forming a finesse $\mathcal{F}\approx2$ cavity. Using direct photodetection in transmission, we resolve the relative displacement of the membranes with a shot-noise-limited imprecision of 7 fm/$\sqrt{\text{Hz}}$, yielding a thermal-noise-limited acceleration sensitivity of 562 n$g/\sqrt{\text{Hz}}$ over a 1 kHz bandwidth centered on the fundamental trampoline resonance (40 kHz). To illustrate the advantage of radiation pressure stabilization, we cold damp the trampoline to an effective temperature of 4 mK and leverage the reduced energy variance to resolve an applied stochastic acceleration of 50 n$g/\sqrt{\text{Hz}}$ in an integration time of minutes. In the future, we envision a small-scale array of these devices operating in a cryostat to search for fundamental weak forces such as dark matter.
       
\end{abstract}

\maketitle

	Cavity optomechanical (COMS) accelerometers employ a micromechanical oscillator as a test mass and an optical microcavity for displacement-based readout.  Key advantages over micro-electromechanical (MEMS) accelerometers---stemming from the small wavelength of optical fields---include quantum-limited readout, ultrahigh bandwidth; and absolute, traceable calibration \cite{krause2012high,guzman2014high,reschovsky2019high}.  An added feature is radiation pressure back-action, which in principle can enhance sensor performance by leveraging optical stiffening and damping effects. Both have been studied extensively in the field of cavity optomechanics \cite{aspelmeyer2014cavity}; however, the predominant use of stiff, radiofrequency nanomechanical resonators in these studies precluded their application to inertial sensing.
	
	First generation COMS accelerometers used delicate nanofabrication and micro-assembly techniques to monolithically integrate a relatively large, acoustic frequency test mass with an optical microcavity.  Krause \emph{et. al.} \cite{krause2012high} fabricated a millimeter-scale Si$_3$N$_4$ nanobeam in the near field of a 1-D photonic crystal (PtC) cavity, and demonstrated a shot-noise-limited acceleration sensitivity of 10 $\mu g/\sqrt{\t{Hz}}$ over a bandwidth of 25 kHz.  Guzman \emph{et. al.} \cite{guzman2014high} achieved similar performance by fixing a fiber cavity to a centimeter-scale silica test mass. More recently, Zhou \emph{et. al.} \cite{zhou2021broadband} formed a flip-chip COMS accelerometer by sandwiching a micro-mirror against a mass-loaded Si$_3$N$_4$ membrane, enabling sub-$\mu$g sensitivity over several kHz, limited entirely by thermal motion of the test mass.
	
	Here we explore a new platform for COMS accelerometery based on a pair of vertically integrated, \emph{unloaded} Si$_3$N$_4$ membranes with different stiffnesses, forming an optical cavity.  Devices based on this  platform are exceptionally easy to fabricate and give access to a panoply of tools developed for Si$_3$N$_4$ membranes over the last decade, including frequency tuning \cite{st2019swept}, PtC patterning (to increase the membrane's reflectivity) \cite{norte2016mechanical}, and access to ultrahigh-$Q$ flexural modes via mode-shape and strain engineering \cite{tsaturyan2017ultracoherent,reinhardt2016ultralow}.  Using a simple trampoline-on-membrane (``TOM'') design as an illustration, we demonstrate a $f\sim 10$ kHz, $m\sim 10$ ng test mass with a $Q$-$m$ product of milligrams.  In this unique parameter regime, the test mass exhibits a sub-$\mu g$ thermal acceleration and at the same time is highly sensitive to radiation pressure.  We exploit this feature to cold damp the test mass to several millikelvin, and show how this cooling can be used to detect $\sim 10\,\t{n}g/\sqrt{\t{Hz}}$ incoherent accelerations.
	
	\begin{figure}[t!]
	\vspace{0mm}
	\centering
	\includegraphics[width=0.4\textwidth]{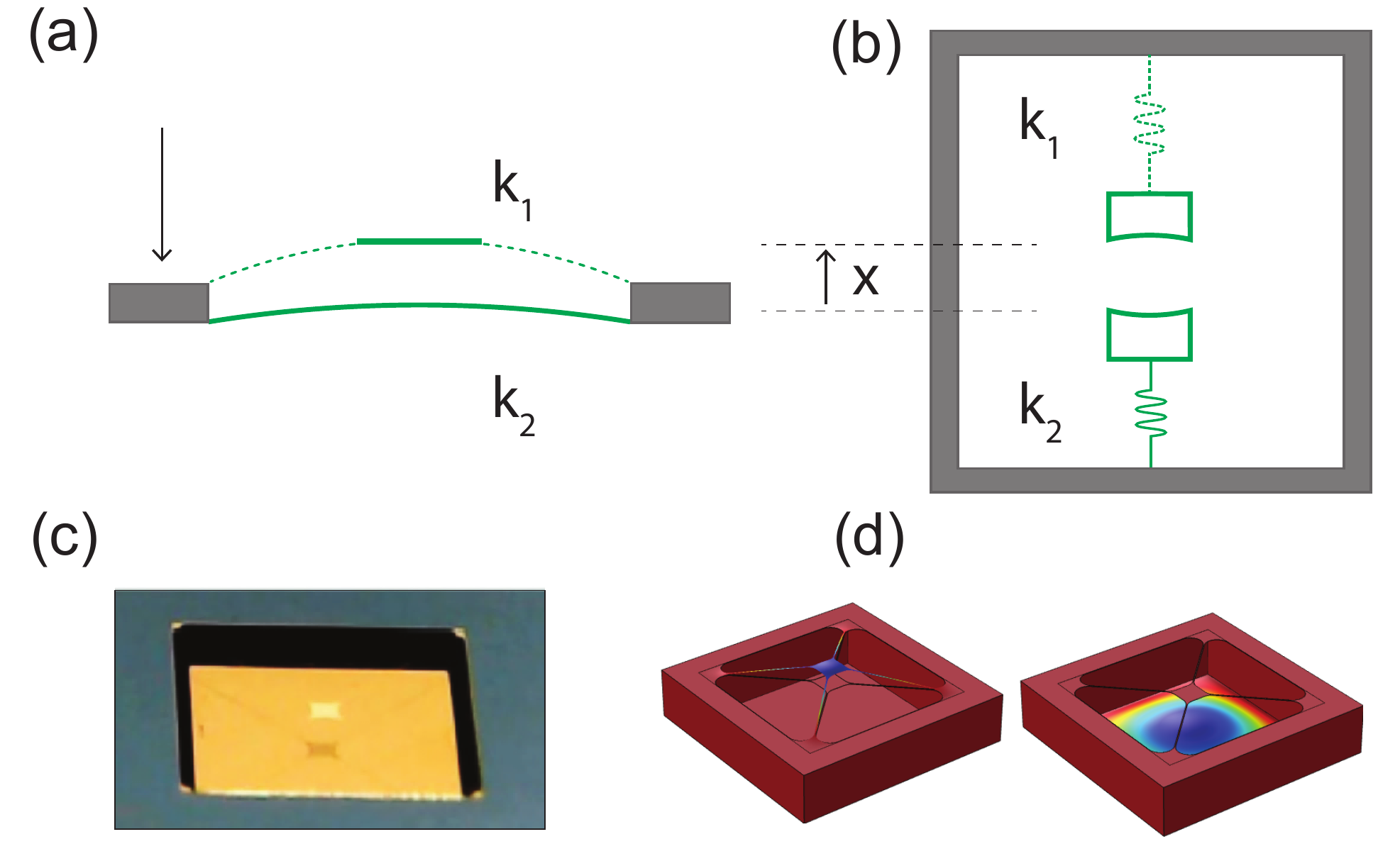}
	\caption{\textbf{Concept for dual-membrane accelerometer:} (a) A pair of membranes with different stiffnesses attached to a common base respond differently to base acceleration. (b) Each membrane is equivalent to a spring-mass system suspended from a common frame. The membranes also act as Fabry-Pérot end-mirrors. (c) Photograph of a trampoline-on-membrane (TOM) accelerometer. (d) Finite-element simulation of the fundamental flexural modes of TOM.}
	\vspace{-4mm}
\end{figure}

 We note that, besides the study of radiation pressure enhanced sensing, development of this platform is motivated by recent proposals to search for weak stochastic accelerations in the acoustic frequency band due to fundamental phenomena such as spontaneous wavefunction collapse and ultralight dark matter \cite{manley2021searching,carney2021mechanical}. These searches can benefit from small-scale arrays of cryogenic optomechanical accelerometers, which call for a simplified approach relative to current technology. 

\begin{figure*}[ht!]
	\vspace{-2mm}
	\centering
	\includegraphics[width=1.9\columnwidth]{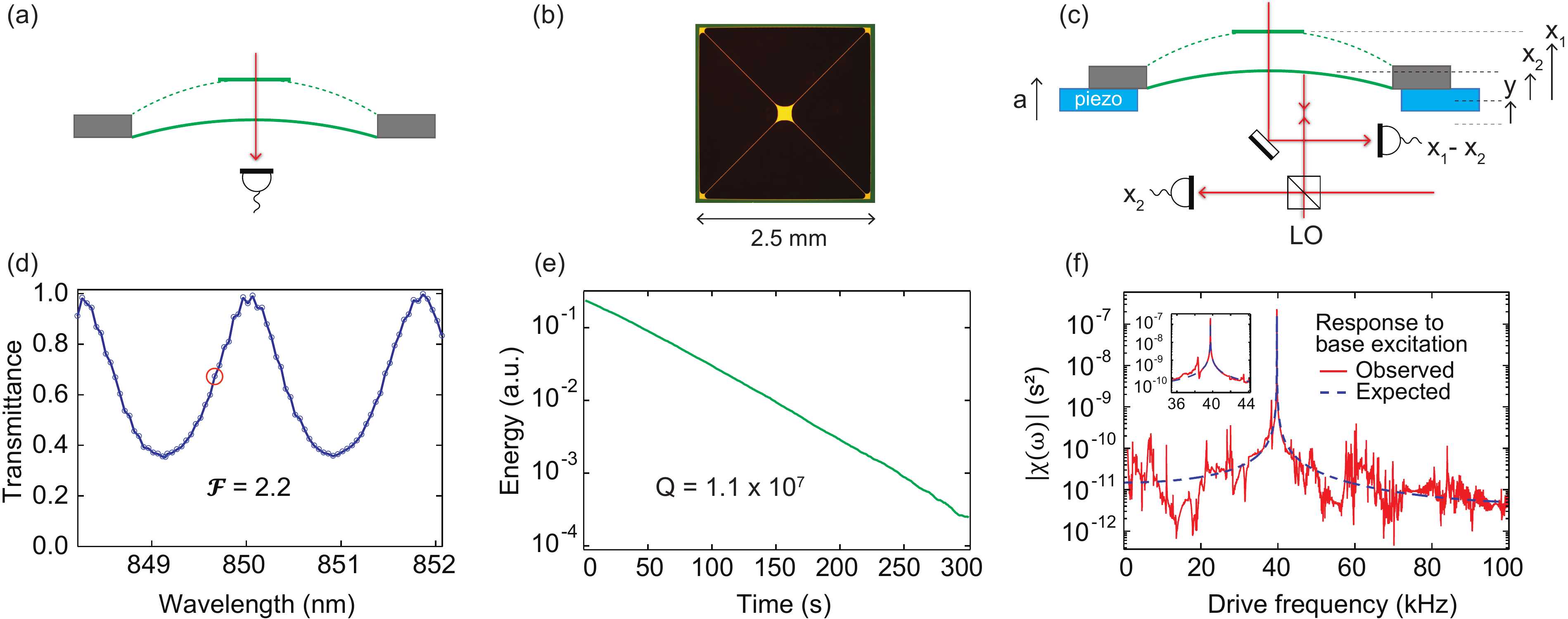}
	\caption{\textbf{Characterization of the dual-membrane accelerometer.} (a), (d) Optical transmission versus wavelength for a laser scanned near 850 nm.  Fitting the interference fringe to an Airy function yields a finesse of 2.2.  Readout is performed near the fringe midpoint, highlighed by a  red circle. (b) Optical micrograph of the trampoline membrane and (e) energy ringdown of its fundamental flexural mode.  Fitting to an exponential yields a $Q$ of $1.1\times10^7$. (c) Scheme for characterizing the response of the dual-membrane accelerometer (see main text) to a base excitation.  (f) Measurement of the response at freqeuencies below the fundamental resonance of the rigid membrane (180 kHz).  Near the 40 kHz trampoline resonance, the response is approximated by the mechanical susceptibility of the trampoline alone.}
	\vspace{-3mm}
\end{figure*}

\emph{Device concept-} The concept behind our approach is illustrated in Fig. 1.  Two dielectric membranes with different stiffnesses are fabricated on opposite sides of the same substrate. (Our membranes are Si$_3$N$_4$ and the substrate is Si; however, the concept is agnostic to material.)  The substrate represents the inertial reference frame and the membranes represent a pair of spring-mass systems; their displacement $x = x_1 - x_2$ is related to the substrate acceleration $a$ by the difference susceptibility
\begin{equation} 
\chi(\omega) = \frac{x(\omega)}{a(\omega)}=\chi_1(\omega)-\chi_2(\omega)
\end{equation}
where
\begin{equation} 
\chi_i(\omega) = \frac{x_i(\omega)}{a(\omega)} \approx \frac{\beta_i}{\omega^2-\omega_i^2+i\omega\omega_i/Q_i} 
\end{equation}
is the susceptibility of membrane $i$, $x_i$ is the displacement of membrane $i$ relative to the substrate, $\omega_i$ and $Q_i$ are the resonance frequency and $Q$ factor of membrane $i$, and $\beta_i\sim 1$ is a unitless factor that depends on the shape of the membrane mode 
(for a square and a trampoline membrane, $\beta \approx (4/\pi)^2$ and $1$ respectively \cite{SI}). 

When the membrane frequencies are different $\omega_1\ne\omega_2$, it is straightforward to show that $\chi(\omega)\approx\chi_i(\omega)$ for frequencies sufficiently close to resonance $\omega\approx\omega_i$.  The dual-membrane system can then be modeled as a canonical spring-mass accelerometer; with the measured displacement yielding an apparent acceleration spectral density
\begin{equation}\label{eq:Sa}
S_a(\omega) = |\chi(\omega)|^{-2} S_x^{\t{imp}}+S_a^\t{th},
\end{equation}
where $S_x^\t{imp}$ is the displacement readout imprecision and
\begin{equation}
S_a^\t{th}(\omega) \approx  \frac{4k_B T \omega_i}{\beta_i^2  m_i Q_i}
\end{equation}
is the apparent acceleration of the substrate due to thermal motion of membrane $i$, with effective mass $m_i$.%

\emph{Trampoline-on-membrane (TOM) accelerometer-} To explore the double-membrane accelerometer concept, we fabricated the device shown in Fig. 1c, consisting of a $2.5 \times 2.5\,\t{mm}^2$, 75-nm-thick Si$_3$N$_4$ ``trampoline" suspended opposite a square membrane of similar dimensions on a 0.2 mm thick Si chip (see Methods). Si$_3$N$_4$ trampolines have been studied before as \emph{local}
force sensors \cite{reinhardt2016ultralow}.  Here, instead, the trampoline serves as a test mass for acceleration of the chip, to which the relatively stiff square membrane is rigidly attached. 
The trampoline we study has a 200-$\mu$m-wide pad and 4-$\mu$m-wide tethers, with fillets tailored to optimize the $Q$ of the fundamental trampoline mode \cite{reinhardt2016ultralow,pluchar2020towards}.  For these dimensions, the fundamental resonance frequency of the trampoline is $\omega_1 = 2\pi\cdot 40$ kHz, the fundamental resonance frequency of the underlying square membrane is $\omega_2 = 2\pi\cdot 180$ kHz, and the effective mass and $Q$ factor of the trampoline are $m_1 = 12$ ng and $Q_1 = 1.1\times10^7$, respectively, implying a thermal acceleration sensitivity of 
$\sqrt{S_a^\t{th}} = 0.56\,\mu g/\sqrt{\t{Hz}}$. 

\begin{figure*}[ht!]
	
	\vspace{-2mm}
	\centering
	\includegraphics[width=1.65\columnwidth]{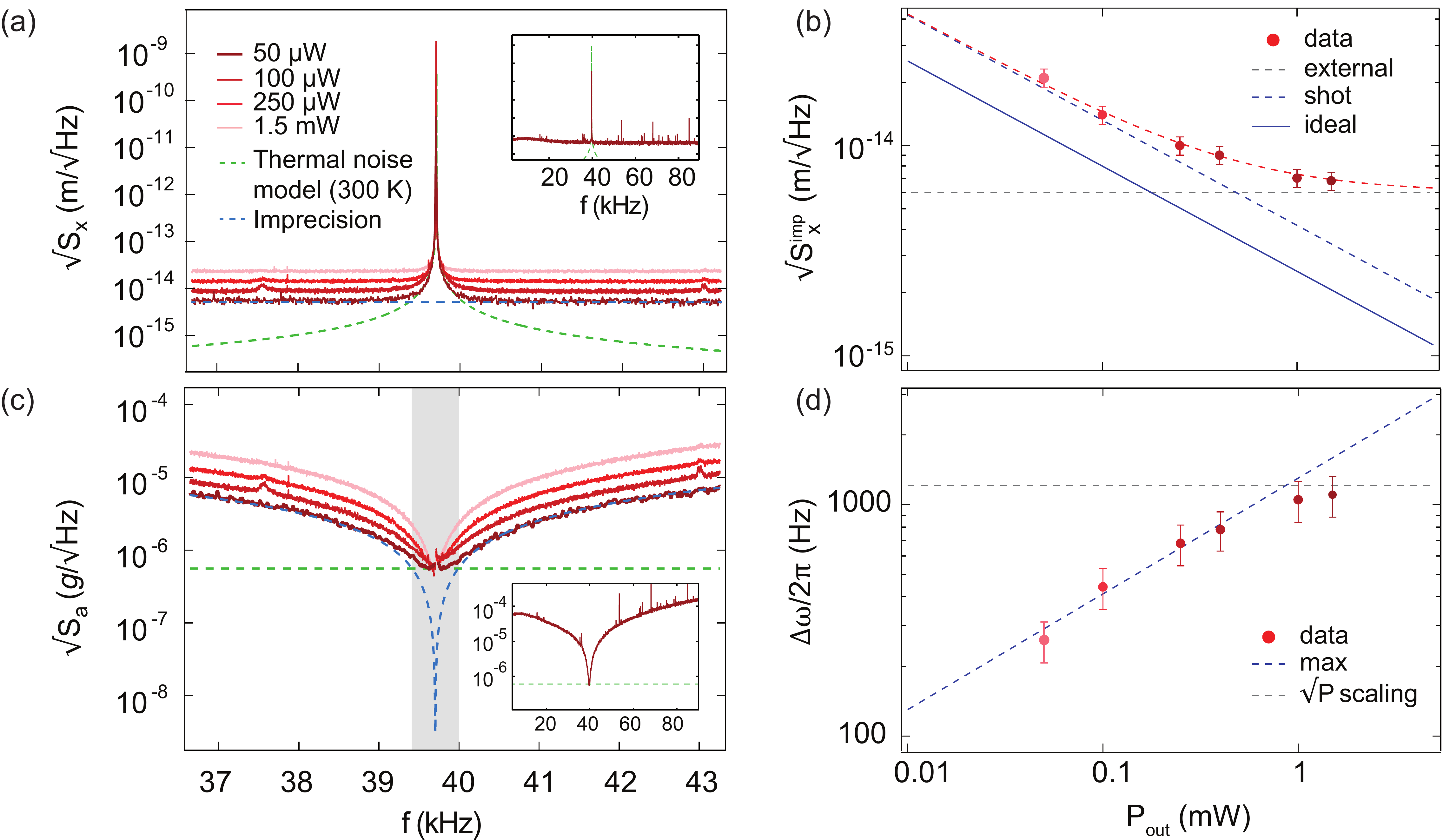}
	\caption{\textbf{Displacement and acceleration sensitivity.} (a) Apparent trampoline-membrane displacement versus output power, calibrated using the thermal noise peak for the fundamental trampoline mode at 40 kHz.  A broadband plot is shown in the inset. (b) Noise floor $S_x^\t{imp}$ of data in panel (a) versus optical power, $P_\t{out}$.  Dashed and solid blue lines are models for shot noise (Eq. 5) with efficiencies of $70\%$ and $100\%$, respectively.  Dashed red line is a model that includes shot noise and an extraneous imprecision noise of $7.0\times 10^{-15}\;\t{m}/\sqrt{\t{Hz}}$ (dashed gray line)  (c) Apparent acceleration of the Si chip, inferred by dividing the data in panel (a) by the acceleration susceptibility (Eqs. 1-3). The gray shaded region indicates the thermal-noise-limited bandwidth $\delta\omega$ for $P_\t{out}=1.5$ mW. (d) Thermal-noise-limited bandwidth versus optical power. Dashed blue and gray lines correspond to models for shot noise (Eq. 6) and a constant extraneous noise (dashed gray line in panel (b)). }
	\vspace{-4mm}
\end{figure*}


As summarized in Fig. 2, we conducted a series of experiments to characterize the performance of the TOM device as an accelerometer.  For these experiments, the device was housed in a high vacuum ($<10^{-7}\,\t{mbar}$) chamber and probed through a viewport with a tunable diode laser (Newport TLB-6716) centered at $\lambda\sim 850$ nm.  The laser was intensity-stabilized using an electro-optic modulator to $-154$ dBc, corresponding to a shot-noise-limited power of 1 mW.  For readout, the transmitted field was directed to a low noise photodetector (Newport PDA36A) and the photocurrent was recorded with a 24-bit digitizer (National Instruments PXI-4462).  

We first studied the performance of the TOM device as an optical cavity, by recording its transmission $\mathcal{T}$ versus laser wavelength $\lambda$ as shown in Fig. 2a.  Comparing to the Airy-function $\mathcal{T}(\lambda)=(1+F\sin^2(4\pi n d/\lambda))^{-1}$, we infer an effective membrane spacing of $d=201\,\mu$m and a finesse coefficient of $F = 4R/(1-R)^2=2$.   These values agree well the specified Si chip thickness and the predicted membrane reflectivity of $R\approx 0.27$ based on a refractive index of $n=2.0$, implying that the dual-membrane etalon behaves like an impedance-matched Fabry-Perot cavity with a finesse $\mathcal{F}\approx \pi\sqrt{F}/2\approx 2.2$. 

We then attempted to characterize the acceleration susceptibility $\chi(\omega)$ of the TOM device, using driven response measurements. 
To record displacement $x$, the laser was tuned to the side of the fringe (red circle in Fig. 2a) and the transmitted power was monitored in real-time.   A piezo located beneath the chip was used to apply a sinusoidal test acceleration.  To estimate the resonant response, a ringdown measurement was performed by transiently exciting the trampoline, yielding $Q=1.1\times 10^7$ as shown in Fig. 1b.  (We note this $Q$ was degraded 4-fold due to dust settling on the sample, but otherwise agrees with dissipation dilution simulations.)  The broadband susceptibility was characterized as shown in Fig. 1c by performing a swept-sine measurement.  To account for structural resonances of the chip and the piezo, for this measurement, the trampoline-membrane displacement $x$ was normalized to an independent homodyne measurement of the square membrane's displacement $x_2$.  We found that the broadband susceptibility agrees qualitatively well with Eq. 2; however, only over a fractional bandwidth of $\sim 20\%$ near the trampoline resonance is it free from spurious features.  As the device was designed especially for resonant sensing, we focus on this region for the remainder of this report.  

In Fig. 3 we present measurements characterizing the acceleration sensitivity of the TOM device, focusing on a 15 kHz wide frequency window centered on the trampoline resonance. We first note that, while low finesse, the high ideality 
of the dual-membrane cavity makes it well suited to quantum-limited displacement readout. 
We confirmed this as shown in Fig. 3a by recording the spectral density of the side-of-fringe photocurrent for several different transmitted optical powers $P_\t{out}$ \cite{SI}.  To calibrate each spectrum in displacement units, the integral under the thermal noise peak at 40 kHz is normalized to $\langle x^2\rangle = k_B T/(m\omega_1^2)$ (photothermal heating was not observed at our operating powers \cite{pluchar2020towards}).
In Fig. 3b, the noise floor is compared to the quantum noise model \cite{SI}
\begin{equation}
    S_x^\t{imp} 
    = \frac{h c\lambda}{6 \eta P_\t{out}\mathcal{F}^2}
\end{equation}
At low powers, $P_\t{out}\lesssim1\,\t{mW}$, the noise floor agrees well with the quantum noise model with an efficiency $\eta \approx 33\%$. At high powers, $P_\t{out}> 1\,\t{mW}$, the noise floor saturates to an extraneous level of
 $\sqrt{S_x^\t{imp}} \approx 7\times 10^{-15}/\sqrt{\t{Hz}}$, consistent with the noise floor of our laser intensity servo.
 

\begin{figure*}[t!]
\vspace{-2mm}
  \centering
  \includegraphics[width=2.0\columnwidth]{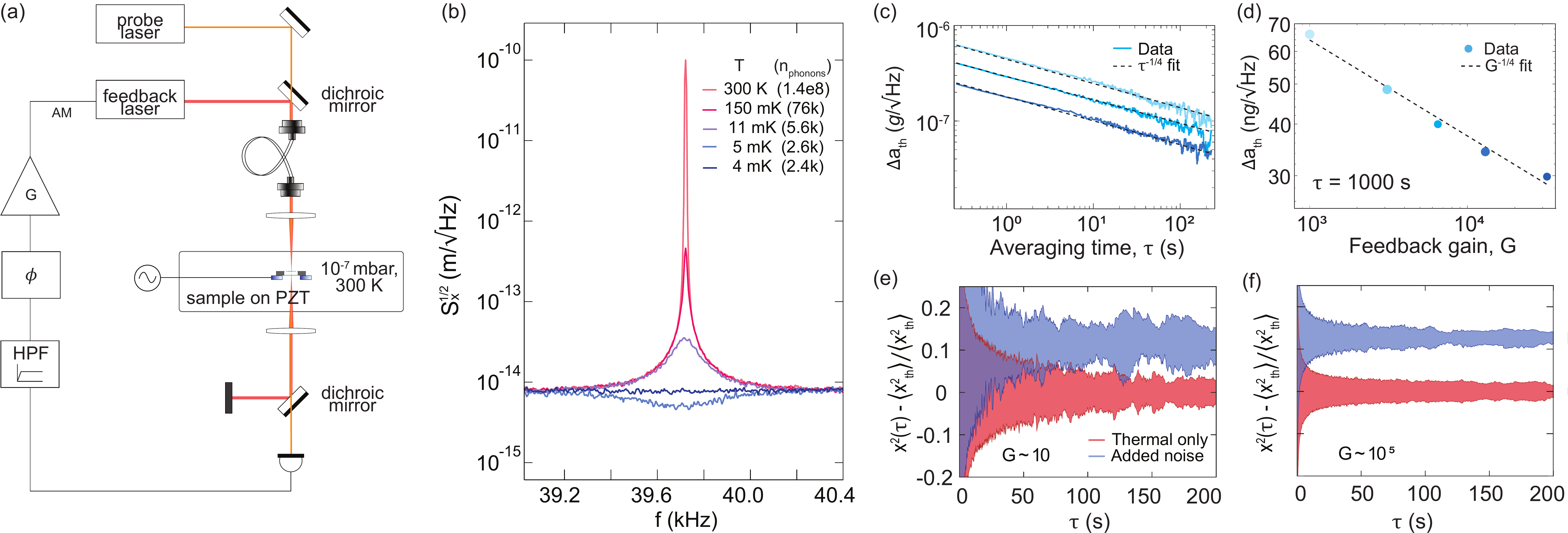}
\caption{\textbf{Radiation pressure enhanced optomechanical accelerometry}. (a) Experimental setup: an 850 nm laser is used to probe the TOM device in transmission using the side-of-the-fringe technique. The processed photosignal---high-pass-filtered (HPF), phase-shifted ($\phi$), and amplified ($g$) ---is used to modulate the intensity of an auxiliary 650 nm laser for radiation pressure feedback. (b) Feedback cooling of the fundamental trampoline mode from room temperature to an effective temperature of 4 mK. (Squashing beyond the noise floor corresponds to an increase in the effective temperature.) (c) Thermal acceleration resolution at 40 kHz as a function of averaging time $\tau$ for several feedback gains. Inset: Displacement spectrum for different $\tau$.  The thermal acceleration resolution is proportional to the variance in the spectrum magnitude.  (d) Thermal acceleration resultion at $\tau = 1000$ s for different feedback gains. (e-f) Acceleration resolution versus time with (blue) and without (red) external white noise applied with a piezo. (e) and (f) correspond to low and high feedback cooling gain, respectively. These plots show that an incoherent acceleration signal can be distinguished from thermal acceleration at much shorter averaging times (f) when the feedback damping (gain) is higher.}
\vspace{-2mm}
\end{figure*}


Chip acceleration spectra inferred from the displacement measurements in Fig. 3a are shown in Fig. 3c.  Each acceleration spectrum is obtained by dividing the displacement spectrum by the model transfer function $|\chi[\omega]|^2$. The observed inverted lineshape has two components: thermal motion of the trampoline, which masquerades as a frequency-independent chip acceleration (solid gray line), and imprecision noise, which has the shape of the inverse susceptibility (dashed gray line). The intersection of these components defines the bandwidth over which the measurement is thermal noise-limited (gray shaded region), and is approximately the product of the mechanical damping rate $\gamma = \omega_1/Q$ and the peak thermal-to-imprecision-noise ratio \cite{SI}
\begin{equation}
\Delta\omega \approx \gamma\sqrt{\frac{S_x^\t{th}(\omega_1)}{S_x^\t{imp}}},
\end{equation}
As shown in Fig. 3e, $\Delta\omega$ increases with power as long as the readout  is shot noise limited, ultimately saturating at $\Delta \omega\approx 2\pi\cdot 1\,\t{kHz}$ due to extraneous noise.  Over this range, we infer a thermal-noise-limited acceleration sensivity of $S_a^\t{th} = 0.6\,\mu g/\sqrt{\t{Hz}}$ based on the trampoline $Q\cdot m$ product.  A moderate increase in the finesse, to $\mathcal{F} = 100$ (e.g. using a PtC membrane \cite{norte2016mechanical}), could in principle extend this sensitivity to baseband ($\Delta\omega \approx \omega_1$).  On the other hand, remarkably, the large resonant thermal motion of the trampoline $S_x^\t{th}(\omega_1)=\tfrac{4k_\text{B} T Q}{m\omega_1^3}=(0.98 \tfrac{\t{nm}}{\sqrt{\t{Hz}}})^2$ implies that thermal noise limited measurements $\Delta\omega > \gamma$ are possible with as little as femtowatts of optical power in the current Fresnel reflection ($\mathcal{F}\sim 1$) arrangement.  

 \emph{Radiation-pressure-enhanced accelerometry- } The preceding results show that a pair of high $Q$ Si$_3$N$_4$ membranes can be used realize an optomechanical accelerometer with sub-$\mu g$ sensitivity over a bandwidth of kilohertz, using milliwatts of optical power.  We now turn our attention to a unique opportunity afforded by the simultaneous high force sensitivity of the membranes, which enables their intrinsic mechanical susceptibility to be overwhelmed by radiation pressure back-action, leading to optical stiffening and damping \cite{aspelmeyer2014cavity}.  Previous work on COM accelerometers cited radiation pressure back-action as a future tool for enhanced sensing \cite{krause2012high}.  For example, optical damping might be used to rectify instabilities and nonlinearities, yielding improved dynamic range; and optical stiffening might be used to increased bandwidth (by frequency scanning) or sensitivity (via loss dilution).  However, the use of relatively large, low $Q$ test masses, and a focus on baseband acceleration sensing, has limited the implementation of these proposals to date.

The TOM device was designed with radiation pressure in mind as a tool for \emph{incoherent} acceleration sensing, inspired by recent proposals to search for a weak inertial force produced by ultralight dark matter \cite{manley2021searching,carney2021mechanical}.  Towards this end, we follow the approach of Gavartin \emph{et. al.} \cite{gavartin2012hybrid} to show that radiation pressure cold-damping (feedback cooling) can be used to reduce the resolving time for an incoherent acceleration measurement. 
The strategy of \cite{gavartin2012hybrid} involves using Bartlett's method to estimate the area $\left< x^2 \right>$ of the thermal noise peak, which is proportional to $S_a^\t{th}$.  The standard deviation of the estimate $\Delta\left< x^2\right>$ determines to the smallest incoherent acceleration that can be resolved atop thermal noise, viz.
\begin{equation}
		S_a^\t{min}(\tau,\gamma_\t{eff}) = \frac{\Delta\left<x^2\right>}{\int_\infty^\infty  \abs{\chi(\omega)}^2 d\omega /2\pi}	\simeq \frac{S_a^\t{th}}{(\gamma_\t{eff}\tau)^{1/2}}
\end{equation}
where $\tau$ is the measurement time, 
and $\gamma_\t{eff}$ and $T_\t{eff}$ are the effective damping rate and temperature of the mechanical mode, respectively, satisfying $\gamma_\t{eff} T_\t{eff} = \gamma T$ for cold damping 
\cite{SI}.  Eqs 6 and 7 imply that $\tau$ can be reduced by as much as the initial signal-to-noise ratio $\sqrt{S_x^\t{th}[\omega_1]/S_x^\t{imp}}$ by cold-damping to the noise floor, $\gamma_\t{eff} = \Delta\omega$.  While this result has been shown to only recover the performance of an optimal Wiener filter estimation strategy \cite{vinante2013dissipative}, it holds a practical advantage in that the Wiener filter must approximate the intrinsic mechanical susceptibility \cite{harris2013minimum}, a task which is difficult to accomplish electronically.

In Fig. 4, we present an experiment demonstrating incoherent acceleration sensing at the level of 50 $\t{n}g/\sqrt{\t{Hz}}$ by cold-damping the fundamental trampoline mode of the TOM device.  The radiation pressure damping force, $F_\text{RP}=-G\gamma_0\dot{x}$, is supplied by a secondary $\lambda\sim 650$ nm laser whose amplitude is modulated with a phase-shifted copy of the photosignal using a pre-amplifier (Stanford Research Systems SR560) and a delay line. The increased net damping, $\gamma=(1+G)\gamma_0$, reduces the effective temperature of the flexural mode from room temperature $T_0 = 300$ K (i.e., $\langle n\rangle = k_\text{B}T_0/\hbar\omega_1\approx1.4\times10^8$ thermal phonons) to an effective temperature of \cite{poggio2007feedback}
\begin{equation}
T = \frac{1}{1+G}T_0+ \frac{G^2}{(1+G)}T_\t{imp}\ge 2\sqrt{T_0 T_\t{imp}}
\end{equation}
where $T_\text{imp}= T_0 \tfrac{S_x^\t{imp}}{S_x^\t{th}(\omega_0)}$ is the effective measurement  temperature \cite{poggio2007feedback}. As shown in Fig. 4b, $S_x^\text{imp}=7\text{ fm}/\sqrt{\text{Hz}}$ allows us to cool the mode to $T\approx$ 4 mK ($\langle n \rangle \approx 2.4\times 10^3$) using an optimal gain of $G = 1.4\times 10^5$. 
Applying the area estimation method for a variety of intermediate gains, we confirmed that $\sqrt{S_a^\t{min}}\propto (G\tau)^{-1/4}$ as shown in Figs. 4c and 4d.  At the optimal gain setting, a apparent acceleration resolution of \textcolor{black}{$\sqrt{S_a^\t{min}}\approx50\,\t{n}g/\sqrt{\text{Hz}}$} was achieved with a measurement time of 200 seconds, limited by the onset of drift in experimental controls \cite{SI}.

Finally, as a demonstration, we used a piezo to apply a \textcolor{black}{$\sim 180\,\t{n}g/\sqrt{\t{Hz}}$ incoherent acceleration to the chip, smaller than the thermal acceleration by a factor of $\sim 3$}. As shown in Figs. 4e and 4f, the area averaging technique was used to estimate the total noise with feedback gains of \textcolor{black}{$G \sim 10$} and $G \sim 10^5$, respectively. Shaded regions highlight the uncertainty of the estimate. For lower gain (Fig. 4e), it takes visibly longer ($\sim 100\,\t{s}$) to resolve the external acceleration noise from thermal noise. For higher gain (Fig. 4f), the resolving time is significantly smaller ($\sim 1\,\t{s}$); however, the signal-to-noise (the vertical distance between the red and blue data) remains unchanged since the feedback is linear.

\emph{Summary and outlook: }
We have presented a platform for optomechanical accelerometry based on a pair of vertically integrated Si$_3$N$_4$ membranes with different stiffnesses, forming an optical cavity. Devices based on this platform are simple to fabricate, can achieve sub-$\mu g$ sensitivity over a bandwdith of kilohertz, and can actively controlled used radiation pressure feedback. As a proof-of-principle, we studied a ``trampoline-on-membrane" device in which a 40 kHz nanotrampoline with a $Q$-$m$ product of 100 mg was integrated opposite a 150 kHz square membrane on a common 200 $\mu$m Si chip, forming a finesse $\mathcal{F}\approx 2$ cavity. Despite its low finesse, the high ideality of the cavity enabled shot-noise-limited readout of the membrane separation at the level of $\t{fm}/\sqrt{\text{Hz}}$, yielding sensitivity to sub-$\mu g/\sqrt{\text{Hz}}$ chip accelerations over a bandwidth of 1 kHz centered at the trampoline resonance, limited by thermal noise.  Radiation pressure feedback was then used to cool the trampoline's fundamental vibration to an effective temperature of 4 mK.  While not affecting the acceleration sensitivity, we confirmed that cold-damping enables emulation of an optimal filter, allowing us to resolve externally applied chip acceleratios at the level of $50 \text{ n}g/\sqrt{\text{Hz}}$ in a integration time of 200 sec.  In the future, we envision integrating PtC mirrors \cite{gartner2018integrated} into the membranes to achieve $\mathcal{F}\sim 100$; in principle this will enable a 100-fold reduction in displacement noise, extending thermal-noise-limited acceleration sensitivity to baseband.  Operating the device in a dilution refrigerator would would enable quantum back-action limited acceleration measurements \cite{manley2021searching} and give access to sensitivities relevant for fundamental weak forces tests. For example, fixing the device to a germanium or beryllium base would give access to equivalence-principle violating (material-dependent) accelerations hypothetically produced by dark photon dark matter \cite{manley2021searching}.

\color{black}

\section{Methods}
Fabrication of the TOM device is simple and scalable:  Starting with a double-side polished Si wafer, Si$_3$N$_4$ is deposited on both sides using laser plasma chemical vapor deposition, the trampoline is patterned on one side using photolithography, and the double-membrane structure is released with a single KOH wet-etch step, the square membrane serving as both an etch stop and a turbulence shield (removing the need for critical point drying).  Similar structures have been fabricated with embedded PtC mirrors \cite{gartner2018integrated}, and a straightforward extension to PnC membranes \cite{tsaturyan2017ultracoherent,bao2020hybrid} is conceivable, including arrays of such structures distributed over a wafer.   

\section*{Acknowledgements}

This work is supported by NSF Grant ECCS-1945832. The authors thank Christian Pluchar for help designing the optical readout system and Utkal Pandurangi and Felipe Guzm\'{a}n for useful conversations about the development of the device.  ARA acknowledges support from a Friends of Tucson Optics Endowed Scholarship.

\bibliography{sorsamp,ref,references}
\end{document}


\preprint{APS/123-QED}

\title{Supplementary Information for ``Membrane-based Optomechanical Accelerometry''}

\author{Mitul Dey Chowdhury}    
\author{Aman R. Agrawal}
\author{Dalziel J. Wilson}
\affiliation{Wyant College of Optical Sciences, University of Arizona, Tucson, AZ 85721, USA}

\begin{abstract}
\end{abstract}
\maketitle
\tableofcontents
\section{Displacement sensitivity}
The sensitivity of the optical displacement readout depends on the transduction factor $\partial i/\partial x$ from a small change $x$ in relative displacement between the cavity's mirrors to a corresponding change in the detection photocurrent $i$, which is proportional to the optical power on the detector. In the ideal case, the photocurrent is free of technical noise and limited only by optical shot noise (``quantum noise''). 
In this section, we derive the expected theoretical limits to shot-noise-limited displacement sensitivity for a direct transmission measurement, illustrated in Figure 2a of the main text.

Consider a direct detection measurement with input power $P_\text{in}$, such that the transmitted power incident on the detector is
\begin{equation}
    P_\textnormal{out} = \zeta T(x,\lambda) P_\text{in}
\end{equation}
where $\zeta$ is an overall loss factor due to systematic losses and $T$ is the transmission of the dual membrane etalon. Modeling the etalon as a Fabry-P\'{e}rot cavity gives
\begin{equation}
    T  = \frac{1}{1 + F \sin^2(\delta)}
\end{equation}
where $F$ is the coefficient of cavity finesse, $\delta\equiv2\pi (L+x)/\lambda$ is the round trip intracavity phase shift, $L$ is the nominal membrane separation (including the optical path length of the membranes), $x$ is the relative membrane displacement, and $\lambda$ is the wavelength of the optical field. Fluctuations of $P_\text{out}$ could be either due to displacement fluctuations or intensity shot-noise. The latter is given by the single-sided power spectral density (PSD)
\begin{equation}
   S_{P}^\text{out}(\omega) = \frac{2hc}{\lambda}P_\textnormal{out}
\end{equation}
In the adiabatic limit, the apparent displacement (``displacement imprecision'') due to shot noise is therefore
\begin{equation}
   S_x^\text{imp}(\omega) \approx \left(\frac{d P_\text{out}}{dx}\right)^{-2}  S_{P}^\text{out}(\omega)
\end{equation}
where 
\begin{equation}\label{eq:fringeSensitivity}
    \pdv{P_\textnormal{out}}{x} = \zeta P_\text{in}\pdv{T(x,\lambda)}{x}
\end{equation}
For a given input power, the shot noise equivalent displacement is minimized when the derivative of the transmission fringe is maximum, that is, on the side of the fringe. Accordingly, we choose a wavelength such that $|\pdv{T}{\lambda}|$ and thus, also, $|\pdv{T}{x}|\propto |\pdv{T}{\lambda}|$ is maximized.

To simplify, Eq. \ref{eq:fringeSensitivity} can be re-expressed in terms of the output power
\begin{equation}
    \pdv{P_\textnormal{out}}{x}=\frac{P_\textnormal{out}}{T}\pdv{T}{x},
\end{equation}

For $F\gg1$ and $\delta\ll 1$ (near-resonance), the transduction factor $|\pdv{T}{x}|$ is maximized for $\delta =\pm 1/\sqrt{3F}$, leading to $T|\pdv{T}{x}|^{-1}=(\lambda/\pi)/\sqrt{3F}$. The resulting expression for the displacement imprecision is
\begin{equation}
     S_x^\textnormal{imp} \approx \frac{4\hbar c \lambda}{3\pi P_\textnormal{out} F} \approx \frac{h c\lambda}{6 P_\textnormal{out}\mathcal{F}^2},
\end{equation}
where we have substituted $F$ with the finesse $\mathcal{F}=\frac{\pi}{2}\sqrt{F}$. (We note that Eq. 7 is valid to within $10\%$ for $\mathcal{F}>2$.)  Finally, we note that $P_\textnormal{out}$ must account for detection inefficiency; this may be incorporated explicitly by letting $P_\textnormal{out}\rightarrow\eta P_\textnormal{out}$, with $0\leq\eta\leq 1$.

\section{Sensitivity bandwidth}
For measurements shown in Fig. 3 of the main text, the acceleration sensitivity is highest at Fourier frequencies near mechanical resonance, where thermal noise overwhelms imprecision noise. The sensitivity bandwidth $\Delta\omega$ is defined as the range of frequencies over which the total acceleration noise is within 3 dB of the mininum (on-resonance) noise.  Here we formalize this concept.

Following Eq. 1-3 of the main text, we first define the noise-equivalent chip acceleration as
\begin{equation}\label{eq:Sa}
    S_a(\omega) = S_a^\text{th}(\omega) + \beta^2[(\omega_0^2 - \omega^2)^2 + (\omega\gamma)^2]S_x^\text{imp}
\end{equation}
where $\beta$ is a modal participation factor described in the main text and defined below. 

On resonance, the total acceleration noise is minimized, attaining a value
\begin{equation}
    S_a(\omega_0) = S_a^\text{th} + (\omega_0\gamma)^2 S_x^\text{imp}.
\end{equation}
The sensitivity bandwidth $\Delta\omega$ is defined such that
\begin{equation}\label{eq:bandwidth}
    S_a(\omega_0+\Delta\omega/2)\approx 2S_{a}(\omega_0)
\end{equation}
Combining Eq. \ref{eq:Sa} and \ref{eq:bandwidth} and assuming $\Delta\omega\ll\omega_0$ gives \begin{equation}\begin{split}
    & S_a^\text{th}(\omega_0) + [(\omega_0\Delta\omega)^2 + (\omega_0\gamma)^2]S_x^\text{imp}\\
    &= 2S_a^\text{th}(\omega_0) + 2(\omega_0\gamma)^2 S_x^\text{imp}.
\end{split}
\end{equation}
yielding
\begin{subequations}\label{eq:bandwidth2}
    \begin{align}
        \Delta\omega&\approx
        \gamma\sqrt{\left(\frac{S_a^\text{th}}{S_x^\text{imp}}\right)\left(\frac{1}{\omega_0\gamma}\right)^2 + 1}\\
        &=\gamma\sqrt{\left(\frac{S_x^\text{th}(\omega_0)}{S_x^\text{imp}}\right)+1}\\
        &\approx\gamma\sqrt{\left(\frac{S_x^\text{th}(\omega_0)}{S_x^\text{imp}}\right)}.
    \end{align}
\end{subequations}
The final approximation assumes the resonant thermal noise is much larger than the imprecision noise, which is valid for our high-$Q$, room-temperature nanomembrane.

We note that Eq. \ref{eq:bandwidth2}c implies that the sensitivity-bandwidth product $\sqrt{S_x^\text{imp}}\Delta\omega$ is power-independent, as observed in Figs. 3b and 3d.


\section{Membrane Driven by an acceleration field}\label{sec:membraneaccelerometer}

\subsection{Modal participation factor}
For accelerometers utilizing a rigid-body test mass (a 1D mass on a spring), base acceleration $a(t)$ leads to a spring compression $x$ following the equation of motion.
\begin{equation}
    \ddot{x}+\gamma\dot{x}+\omega_0^2x = a
\end{equation}

By contrast, deformation of a membrane in response to base acceleration depends on its mode-shape.  To see this, consider a membrane in the $y$-$z$ plane subject to an acceleration field $a(y,z,t)$ in the $x$ direction.  This is equivalent to driving each differential mass element in the membrane with a force $dF = adm = a(x,y)\rho h dxdy$, where $\rho$ is the mass density and $h$ is the membrane thickness. Neglecting dissipation, the membrane displacement profile $u(x,y,t)$ then obeys the equation of motion
\begin{equation}\label{eq:1}
    \rho\ddot{u} + \sigma\nabla^2 u = \rho a
\end{equation}
where $\sigma$ is the biaxial stress. 

Eq. \ref{eq:1} can be solved by first considering homogeneous ($a = 0$) solutions of the form
\begin{subequations}\begin{align}
    u(x,y,t) &= \phi(x,y)\eta(t)\\
    \nabla^2\phi + k^2\phi &= 0\\
    \ddot{\eta} + \omega^2 \eta &= 0\\
    \omega^2 + (\sigma/\rho)k^2 &= 0
\end{align}\end{subequations}
These are vibrational modes with shapes $\phi(x,y)$ and frequencies $\omega$.  For a square membrane of width $L$, e.g.
\begin{subequations}\begin{align}
    \phi_{ij}(x,y) &=\sin\left(\frac{i\pi x}{L}\right)\sin\left(\frac{j\pi y}{L}\right)\\
    k_{ij} & = \frac{\pi}{L}\sqrt{i^2+j^2} \\
    \omega_{ij} &=  \frac{\pi}{L}\sqrt{\frac{2\sigma}{\rho}}\sqrt{\frac{i^2+j^2}{2}}
\end{align}\end{subequations}
Now assume a general solution to \eqref{eq:1} of the form
\begin{equation}
    u(x,y,t) = \sum_{ij}\phi_{ij}(x,y)\eta_{ij}(t)
\end{equation}
Invoking orthoganility of the mode shapes
\begin{equation}
    \int\phi_{ij}\phi_{i'j'}dxdy \propto \delta_{i,i'}\delta_{j,j'}
\end{equation}
gives 
\begin{equation}\label{eq:6}
    \ddot{\eta}_{ij}(t)+\omega_{ij}^2\eta_{ij}(t)= a_{ij}(t)
\end{equation}
where
\begin{equation}
    a_{ij}(t)= \frac{\int \phi_{ij}(x,y)a(x,y,t)dxdy}{\int \phi^2_{ij}(x,y)dxdy}
\end{equation}
is the generalized acceleration of mode $ij$.

For an acceleration field with a constant shape $a(x,y,t) = \phi_a(x,y)a(t)$, 
Eq. \ref{eq:6} simplifies to
\begin{equation}\label{eq:7}
    \ddot{\eta}_{ij}(t)+\omega_{ij}^2\eta_{ij}(t)= \beta_{ij} a(t)
\end{equation}
where 
\begin{equation}\label{eq:beta}
\beta_{ij} = \frac{\int \phi_{ij}(x,y) \phi_a(x,y) dxdy}{\int \phi^2_{ij}(x,y)dxdy}
\end{equation}
is the modal ``participation'' factor.

Base acceleration corresponds to a uniform ($\phi_a = 1$) acceleration field.  In this case, for a square membrane
\begin{equation}
\beta^\text{sq}_{ij} = \begin{cases}
      \left(\frac{4}{\pi}\right)^2\frac{1}{ij} & i \wedge j \in {1,3...}\\
      0 & i \vee j \in {2,4...}
    \end{cases}
\end{equation}

For the fundamental mode of trampoline with vanishingly narrow tethers and a pad width much smaller than the tether length (approximating our device, shown in Fig. 2b), Eq. \ref{eq:beta} implies $\beta\approx 1$.

\subsection{Thermal noise}

Thermal motion can be modeled by adding ad-hoc damping and Langevin terms to \eqref{eq:6}, viz. 
\begin{equation}\label{eq:6langevin}
    \ddot{\eta}_{ij}+\gamma_{ij}\dot{\eta}_{ij}+\omega_{ij}^2\eta_{ij}= a_{ij} + a^\text{th}_{ij}
\end{equation}
where
\begin{subequations}\label{eq:effmass}\begin{align}
    a_{ij}^\text{th}& = \sqrt{2 k_B T\gamma_{ij}/m_{ij}}\xi(t)\\
    \langle \xi (t)\rangle &= 0\\
    \langle \xi(t)\xi(t+\tau)\rangle &= \delta(\tau)\\
    m_{ij} &= \int\rho\phi_{ij}^2 hdxdy.
\end{align}\end{subequations}
Here $\gamma_{ij}$ is a phenomenological damping rate and $m_{ij}$ is an effective mass defined such that $\langle \eta_{ij}(t)^2\rangle = k_B T/m_{ij}\omega_{ij}^2$. For a square membrane, for example,  
\begin{equation}
    m^\text{sq}_{ij} = \frac{\rho h L^2}{4} = \frac{m_\text{phys}}{4}
\end{equation}
where $m_\text{phys}$ is the physical mass.  

For the fundamental mode of trampoline with vanishingly narrow tethers and a pad width much smaller than the tether length (approximating our device, shown in Fig. 2b), Eq. \ref{eq:effmass}d implies $m\approx m_\text{phys}$.

\subsection{Effective displacement}

In practice we measure an effective displacement
\begin{equation}
x(t) = \int \phi_x(x,y) u(x,y,t) dxdy
\end{equation}
where $\phi_x(x,y)$ is the point-spread function of the measurement apparatus.

Modal decomposition gives
\begin{subequations}\label{eq:xeff}\begin{align}
    x(t) & = \sum_{ij}\eta_{ij}(t)\int\psi_x(x,y)\phi_{ij}(x,y)dxdy\\
    & = \sum_{ij}\alpha_{ij}\eta_{ij}(t)\\
    & = \sum_{ij}x_{ij}(t)
\end{align}\end{subequations}
with $x_{ij}$ representing the effective displacement of each mode.   The equation of motion for $x_{ij}$ is 
\begin{equation}\label{eq:6langevindisplacement}
    \ddot{x}_{ij}+\gamma_{ij}\dot{x}_{ij}+\omega_{ij}^2 x_{ij}= \alpha_{ij}\left(a_{ij} + a^\text{th}_{ij}\right)
\end{equation}
which includes an extra overlap factor $\alpha_{ij}$.  

For a homogeneous square membrane probed at its midpoint $\phi_x(x,y) = \delta(x-L/2)\delta(y-L/2)$, e.g.
\begin{equation}
\alpha^\text{sq}_{ij} =\begin{cases}
      1 & i \wedge j \in {1,3...}\\
      0 & i \vee j \in {2,4...}\end{cases}
\end{equation}

For the fundamental mode of trampoline with vanishingly narrow tethers and a pad width much smaller than the tether length (approximating our device, shown in Fig. 2b), Eq. \ref{eq:xeff} implies $\alpha\approx 1$.

\section{Acceleration susceptibility of the dual-membrane system}
In this section, we derive susceptibility of the dual-membrane system to chip acceleration.  Following Sec. \ref{sec:membraneaccelerometer} and Fig. 2c of the main text, we consider the response $x_i$ of a membrane to a displacement $y$ of its base (or chip). We limit our analysis to the fundamental mode with resonance $\omega_i = 2\pi f_i$ and damping rate damping $\gamma_i = \omega_i/Q_i$, where $i\in\{1,2\}$ corresponds to the top ($i=1$) or bottom $(i=2)$ membrane.  Eq. \ref{eq:6langevindisplacement} yields
\begin{equation}
    \Ddot{x}_i(t)  + \gamma_i\dot{x}_i(t) + \omega_i^2 x_i(t) = \beta_i\Ddot{y}(t),
\end{equation}
leading to the transfer function
\begin{equation}
    x_i(\omega) = \frac{\beta_i}{\left(\omega_i^2 - \omega^2\right)+ i\gamma_i\omega}  a(\omega)\equiv \chi_i(\omega)a,
\end{equation}
where $a(\omega)=-\omega^2 y(\omega)$ is the base (chip) acceleration and $\chi_i(\omega)$ is the acceleration susceptibility of each membrane. The relative susceptibility
\begin{equation}
    \chi(\omega)=\frac{x_1(\omega)-x_2(\omega)}{a(\omega)} \equiv  \frac{x(\omega)}{a(\omega)} = \chi_1(\omega)-\chi_2(\omega)
\end{equation}
determines the differential displacement of the dual-membrane system when the base is accelerated. 

The various acceleration susceptibilities for our "trampoline on membrane'' (TOM) device are plotted in Figures 1 and 2. Note that, since the square membrane ($i=2$) is stiffer than the trampoline ($i=1$), then $\chi(\omega\ll\omega_2)\approx\chi_1(\omega)$.

\begin{figure}[h!]\label{fig:xbyy}
  \centering
  \includegraphics[width=0.45\textwidth]{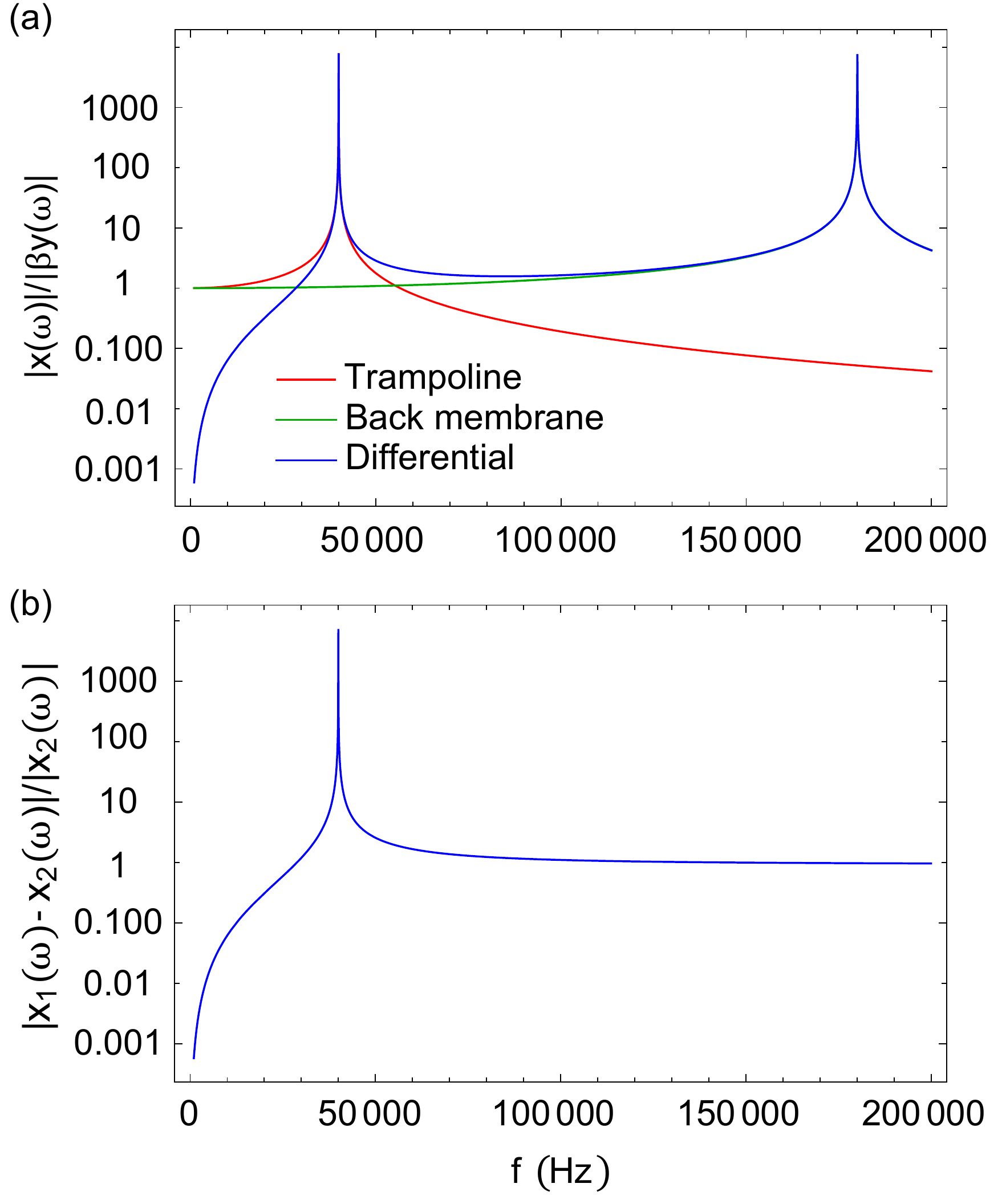}
\caption{Predicted response of individual membranes $x_{1,2}$ and the combined system ($x = x_1 - x_2$) to base displacement $y$. (a) Susceptibility to base displacement vs frequency (e.g. $x(\omega)/y(\omega)  = \omega^2|\chi(\omega)|$). (b) Ratio of relative displacement to the displacement of the square membrane, using Eq. \ref{eq:x1x2overx2}.}
\end{figure}
\begin{figure}[h!]
  \centering
  \includegraphics[width=0.45\textwidth]{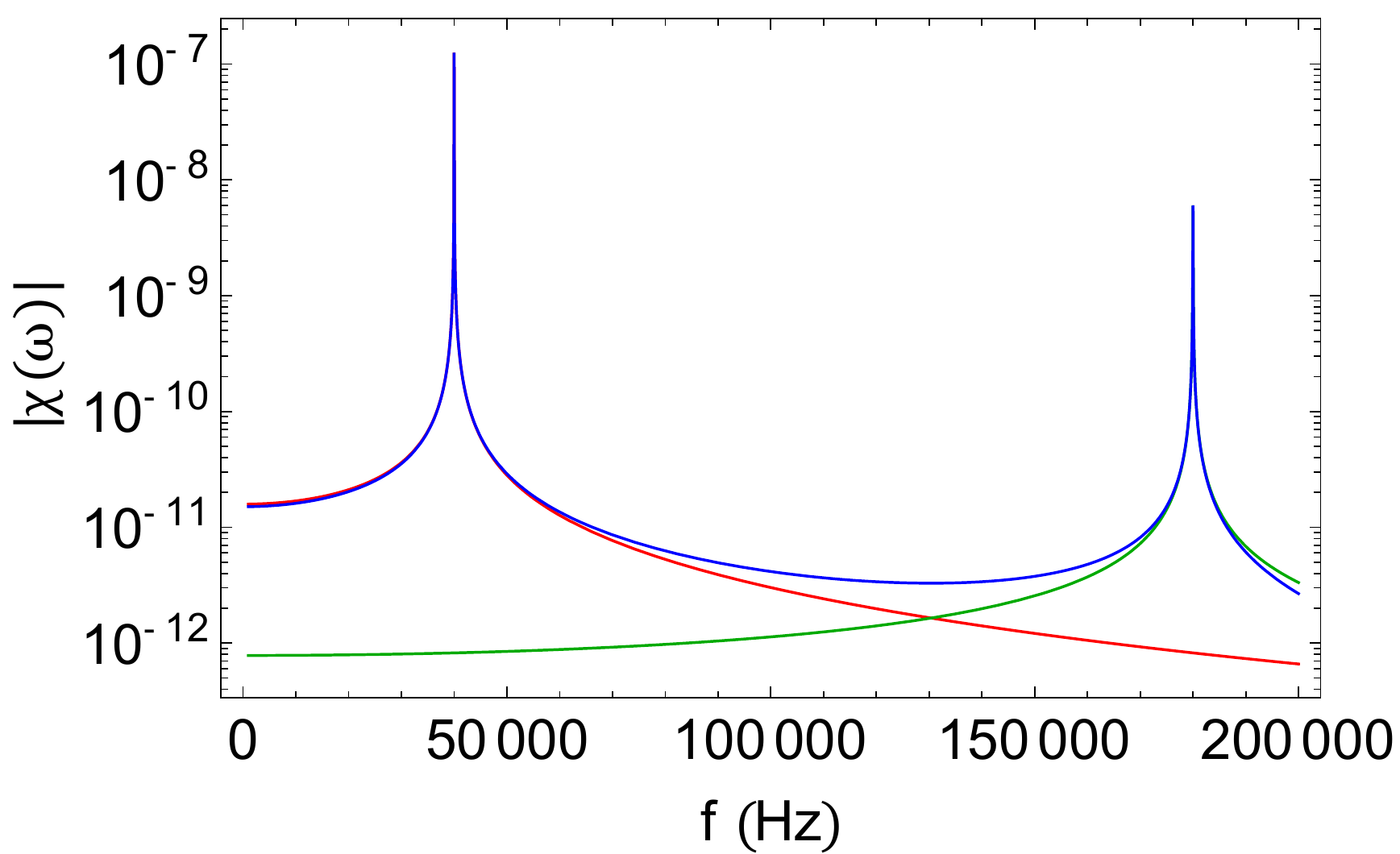}
\caption{Acceleration susceptibilities of the individual membranes $\chi_{i}(\omega)$ and the combined system $\chi(\omega)$. Red, green, and blue correspond to the trampoline ($i=1$), square membrane ($i=2$), and combined system, respectively.}
\end{figure}

As shown in Fig. 2f of the main text, we characterized the acceleration susceptibility of the TOM device by subjecting the chip to sinusoidal displacement with a piezo. The drive frequency was swept up to 100 kHz while simultaneously monitoring the relative membrane displacement $|x_1(\omega)-x_2(\omega) = x(\omega)|$ and back membrane displacement $|x_2(\omega)|$. Their ratio
\begin{equation}\label{eq:x1x2overx2}
    \frac{|x_1(\omega)-x_2(\omega)|}{|x_2(\omega)|}\approx\frac{|x(\omega)|}{\beta_2|y(\omega)|}=\frac{\omega^2}{\beta_2}|\chi(\omega)|
\end{equation}
can be used to infer $|\chi(\omega)|$, as shown in Figure 2f of the main text. Above, we have used the approximation $|x_2(\omega)|\approx \beta|y(\omega)|$ for $\omega\ll\omega_2$.

\section{Resolution of thermal acceleration}
In this section, we derive the uncertainty involved in estimating accelerations by resonant detection. In essence, the ``resolution'' of the measurement is determined by the variance in the energy estimate obtained by integrating the area under the resonance peak \cite{gavartin2012hybrid}.

Consider a harmonic oscillator of effective mass $m$, stiffness $k$, resonance frequency $\w_0=\sqrt{k/m}$, and viscous damping $\g$, subject to stochastic thermal acceleration noise $a_\textnormal{th}=\sqrt{2 k_\textnormal{B} T\gamma/m}\xi(t)$:
\begin{equation}\label{eq:Langevin}
    \Ddot{x} + \g \dot{x} + \wo^2 x = a_\textnormal{th}.
\end{equation}
In thermally limited displacement-based accelerometers, the smallest resolvable acceleration is set by the imprecision in estimating the thermal noise.

The thermal energy $U=x^2$ can be estimated from the oscillator's (double-sided) power spectral density (PSD), \begin{equation}\label{eq:PSD}
    S_x(\omega) =\left(2 k_\text{B} T \g/m\right) \left[(\wo^2-\w^2)^2 + (\g\w)^2\right]^{-1}
\end{equation} 
from the area under the thermally driven resonance peak gives the average potential energy imparted by the thermal process 
\begin{equation}\label{eq:averageU}
    \x{U} = \x{x^2} = 2\pi\int_{-\infty}^\infty d\omega S_x(\omega) = \frac{k_\text{B} T}{m\wo^2}.
\end{equation}
The final equality derives from the equipartition of energy in thermal equilibrium. Note that $x$ is a zero-mean Gaussian with variance $\x{x^2}$, which makes $U$ a $\chi^2$ distribution. The variance of $U$, $\x{\D U}$, is the 4th central moment of $x$, which may be calculated using the standard method for Gaussian probability densities. Alternately, using thermodynamic identities,
\begin{equation}\label{eq:deltaU}
    \x{\D U^2} = -  \pdv{\x{U}}{\beta_T} = 2 \x{U}^2,
\end{equation}
where $\beta_T=\frac{m\wo^2}{2}\frac{1}{k_\text{B} T}$.
The goal is to reduce this variance by increasing the sample-size of the energy estimate -- the PSD-area. In practice, the PSD is approximated by a periodogram estimate, obtained by averaging the displacement read-out signal $x(t)$ over a finite integration time, $\tau$. The resulting energy estimate is,
\begin{equation}
    \e{U}(\tau) = \frac{1}{\t}\int_0^\t dt x^2(t).
\end{equation}
Note that the estimator $\e{U}(\tau)$ is unbiased, since
\begin{equation}
    \x{\e{U}(\t)} = \x{\frac{1}{\t}\int_0^\t dt x^2(t)} = \frac{1}{\t}\int_0^\t dt \x{x^2(t)} = \x{x^2},
\end{equation}
where we have used ergodicity to exchange the ensemble and time-averages (second equality) and to assert that the variance of $x(t)$ is independent of $t$ in thermal equilibrium (final step). Finally, the imprecision in thermal energy estimation is simply the variance in the estimator, $\s^2\left(\t\right)\equiv\x{\D\e{U}^2\left(\t\right)}$. We find that:
\begin{align}\label{eq:varianceTau}
    \begin{split}
        \s^2(\t) &= \x{\left(\frac{1}{\tau}\int_0^\t dt \D U(t)\right)\left(\frac{1}{\t}\int_0^\t dt' \D U(t')\right)}\\
        &= \frac{1}{\t^2}\x{\int_0^\t \int_0^\t dt dt' \D U(t) \D U (t')}\\
        &= \frac{1}{\t^2} \int_0^\t \int_0^\t dt dt' \x{\D U(t) \D U (t')},
    \end{split}
\end{align}
using the ergodic assumption in the last step. 

Correlations in energy would decay at the damping rate $\g$, given by
\begin{equation}\label{eq:expEAC}
       \c(t,t') = 2 \x{U}^2 e^{-\g \abs{t'-t}},
\end{equation}
which ensures that, at any $t'=t$, $\x{\D U(t) \D U (t')} = \x{\D U^2} = 2\x{U}^2$ which is the variance in the potential energy for Gaussian thermal noise (Eq. \ref{eq:deltaU}), as expected. 

The exponentially decaying autocorrelation (Eq. \ref{eq:expEAC}) in conjunction with  Eq. \ref{eq:varianceTau}, yields the energy imprecision:
\begin{align}\label{eq:energyvar}
    \begin{split}
        \s^2(\t) &= 2\x{U}^2\frac{1}{\t^2} \int_0^\t \int_0^\t dt dt' e^{-\g \abs{t'-t}}\\
        &= 2 \x{U}^2\left[\frac{2 \left(\g \t + e^{-\g \t} -1 \right)}{\left(\g \t \right)^2}\right].
    \end{split}
\end{align}
As a sanity check, consider the limit $\tau\rightarrow 0$: the bracked term $\left[\ldots\right]\rightarrow 1$ on the RHS of Eq. \ref{eq:energyvar}, yielding $\x{\D\e{U}^2(\tau)} \rightarrow 2 \x{U}^2$. This is consistent with the expected energy variance (Eq. \ref{eq:deltaU}), when variance-reduction by averaging is not available (sample size is unity). Finally, we note that Eq. \ref{eq:energyvar} is consistent with the estimation uncertainty reported by Gavartin \emph{et. al.} \cite{gavartin2012hybrid} in the context of feedback-enhanced force estimation. In this work, we have focused on the $\tau\gg1/\gamma$ regime, where $\s^2(\t)\propto(\gamma\t)^{-1/2}$.

\begin{figure}[h]
\label{Fig.1}
\centering
\includegraphics[width=0.4\textwidth]{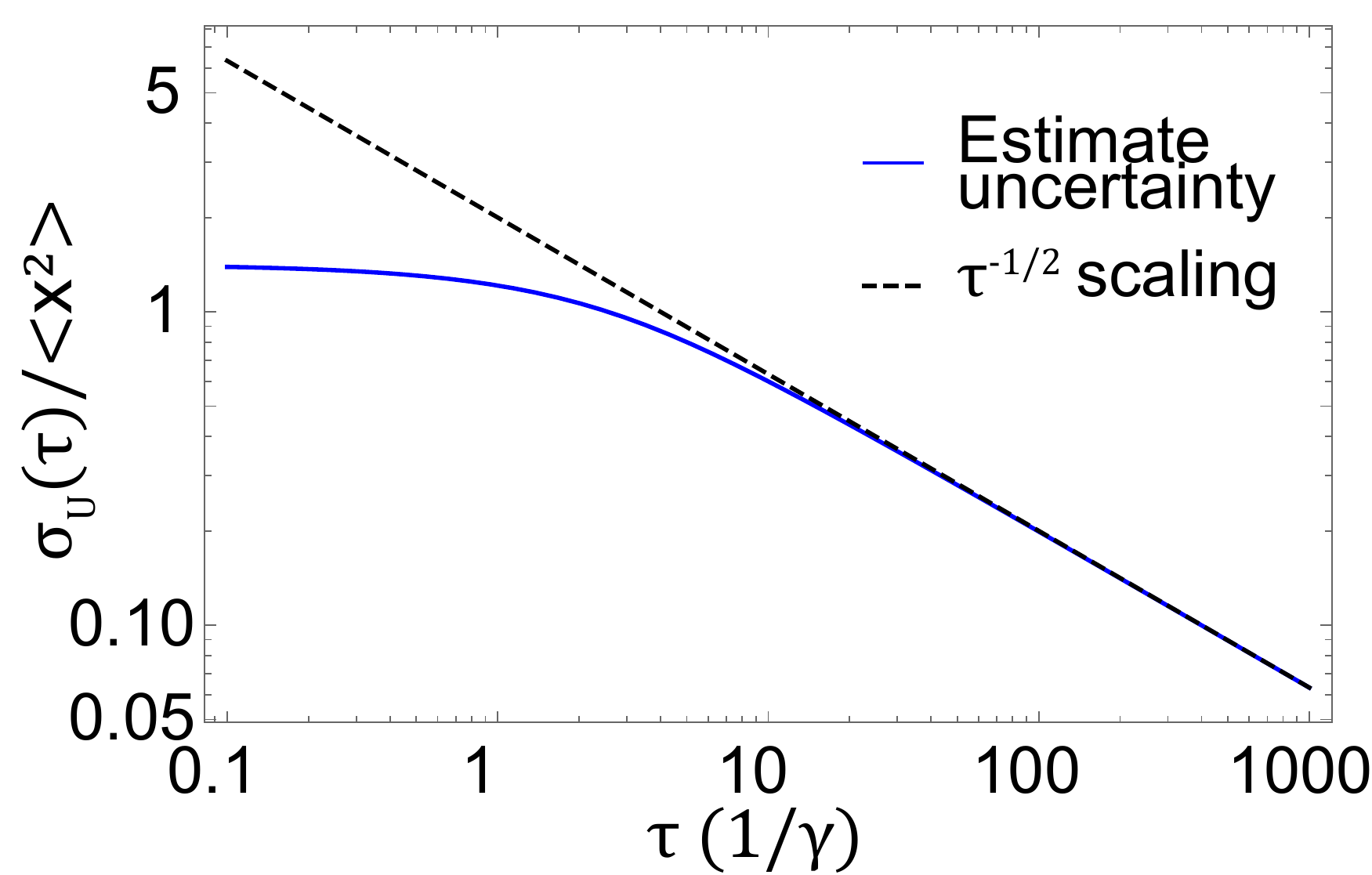}
\caption{Uncertainty in the thermal energy estimate as a function of averaging time for arbitrarily chosen $\gamma$.}
\end{figure}

\section{Limits of time averaging}
While time averaging improves the estimate of incoherent accelerations, long-term instabilities or drifts in the physical experimental parameters and controls limit the time over which averaging is beneficial. To explore the duration over which averaging leads to a reduction in the estimation uncertainty, we consider, as a function of averaging time, the Allan deviation (AD) of the thermal acceleration -- a commonly used technique in the field of frequency metrology. AD obeys the same scaling as standard deviation, until the onset of long-term drift in the parameter of interest. As plotted in Figure 4, reduction in the AD is seen only until $\sim 200 s$. For longer $\tau$, the AD deviates from the expected $\tau^{-1/4}$ scaling, and eventually starts to increase -- telltale of experimental drift.

\begin{figure}[H]\label{fig:adev}
\vspace{5mm}
  \centering
  \includegraphics[width=0.95\columnwidth]{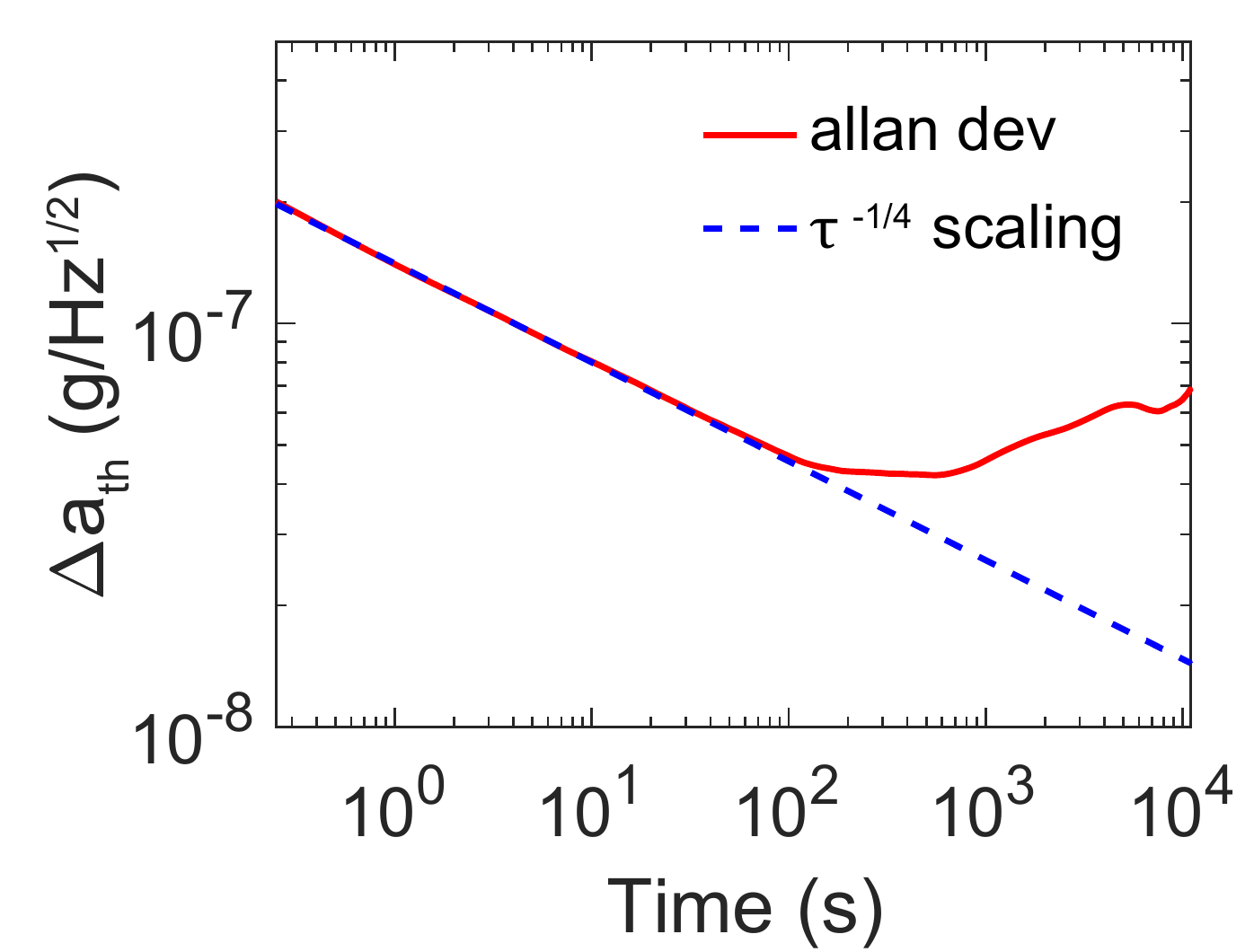}
\caption{Allan deviation of thermal area estimate as a function of averaging time for the optimal damping case. The deviation from $\tau^{-1/4}$ scaling after roughly 200 seconds indicates that averaging further does not improve the estimate.}
\end{figure}
\bibliography{sorsamp}